\documentstyle[12pt]{article}

\def\Dslash{\not{\hspace{-.05in}D}}
\def\kp{K\"ahler potential }

\def\Tr{{\rm Tr}}

\def\L{{\cal L}}

\def\n{\bar{n}}
\def\D{{\cal D}}

\def\bfZ{{\bf Z}}

\def\bt{\bar{t}}

\def\s{\bar{s}}

\def\tf{\tilde{f}}
\def\tF{\widetilde{F}}

\catcode`\@=11

\def\ksection{\arabic{section}}

\def\@normalsize{\@setsize\normalsize{15pt}\xiipt\@xiipt
\abovedisplayskip 14pt plus3pt minus3pt%
\belowdisplayskip \abovedisplayskip
\abovedisplayshortskip  \z@ plus3pt%
\belowdisplayshortskip  7pt plus3.5pt minus0pt}

\def\small{\@setsize\small{13.6pt}\xipt\@xipt
\abovedisplayskip 16pt plus3pt minus3pt%
\belowdisplayskip \abovedisplayskip
\abovedisplayshortskip  \z@ plus3pt%
\belowdisplayshortskip  7pt plus3.5pt minus0pt
\def\@listi{\parsep 4.5pt plus 2pt minus 1pt
            \itemsep \parsep
            \topsep 9pt plus 3pt minus 3pt}}

\def\underline#1{\relax\ifmmode\@@underline#1\else
        $\@@underline{\hbox{#1}}$\relax\fi}
\@twosidetrue

\catcode`@=12

\catcode`\@=11

\def\thesection{\arabic{section}.}

\def\ps@headings{\def\@oddfoot{}\def\@evenfoot{}}
\def\@oddhead{\hbox{}\hfill
        \makebox[.5\textwidth]{\raggedright\ignorespaces --\thepage{}--
        \hfill }}
\def\@oddhead{\hbox{}\hfill --\thepage{}-- \hfill}
\def\@evenhead{\@oddhead}

\ps@headings
\catcode`\@=12

\relax

\newcounter{appendix}
\def\appendix{\par
 \reset{equation}
 \addtocounter{appendix}{1}
 \def\thesection{\Alph{appendix}.}
 \def\ksection{\Alph{appendix}}}

\begin{document}
\newcommand{\dal}{\raisebox{0.085cm}
{\fbox{\rule{0cm}{0.07cm}\,}}}
\newcommand{\ot}{\frac{1}{3}}
\newcommand{\al}{\alpha^{\prime}}
\newcommand{\mst}{M_{\scriptscriptstyle \!S}}
\newcommand{\mpl}{M_{\scriptscriptstyle \!P}}
\newcommand{\dv}{\int{\rm d}^4x\sqrt{g}}
\newcommand{\lv}{\left\langle}
\newcommand{\rv}{\right\rangle}
\newcommand{\ph}{\varphi}
\newcommand{\sbar}{\,\overline{\! S}}
\newcommand{\tbar}{\overline{T}}
\newcommand{\ybar}{\overline{Y}}
\newcommand{\nbar}{\,\overline{\! N}}
\newcommand{\qbar}{\,\overline{\! Q}}
\newcommand{\phb}{\overline{\varphi}}
\newcommand{\cm}{Commun.\ Math.\ Phys.~}
\newcommand{\pr}{Phys.\ Rev.\ D~}
\newcommand{\pl}{Phys.\ Lett.\ B~}
\newcommand{\np}{Nucl.\ Phys.\ B~}
\newcommand{\e}{{\rm e}}
\newcommand{\gsi}{\,\raisebox{-0.13cm}{$\stackrel{\textstyle
>}{\textstyle\sim}$}\,}
\newcommand{\lsi}{\,\raisebox{-0.13cm}{$\stackrel{\textstyle
<}{\textstyle\sim}$}\,}
\newcommand{\lag}{\cal L}
\newcommand{\res}{\langle{\rm Re}S\rangle}
\newcommand{\sre}{S_{\!\scriptscriptstyle R}}
\newcommand{\tre}{T_{\!\scriptscriptstyle R}}
\newcommand{\apr}{\widetilde{A}}
\begin{titlepage}
\vspace*{-2.8cm}
\begin{flushright}
LBL--31464\\[-2mm]
UCB-PTH-91/65\\[-2mm]
NUB--3039\\[0.1 in]
\end{flushright}
\begin{center}
\large\sc
Axion Couplings and Effective Cut-offs\\[-2mm]
in Superstring Compactifications
\end{center}
\par \noindent
\begin{center}
{\large Mary K. Gaillard}\raisebox{1mm}{$^{\,\star}$}\\[-1mm]
{\sl Department of Physics, University of California}\\[-2mm]
{\sl and Physics Division, Lawrence Berkeley Laboratory}\\[-2mm]
{\sl Berkeley, CA 94720, U.S.A.}\\[3mm]
{\large and}\\[5mm]
{\large  T.R. Taylor}\raisebox{1mm}{$^{\,\dag}$}\\[-1mm]
{\sl Department of Physics, Northeastern University}\\[-2mm]
{\sl Boston, MA 02115,
U.S.A.}\\[0.2in]
\end{center}
\par
\begin{center}
{\bf Abstract}\\
\end{center}
\begin{quote}
We use the linear supermultiplet formalism of supergravity to study axion
couplings and chiral anomalies in the context of field-theoretical
Lagrangians
describing orbifold compactifications beyond the classical
approximation. By matching amplitudes computed in the effective
low energy theory with the results of string loop calculations
we determine the appropriate counterterm in this effective theory that
assures modular invariance to all loop order. We use
supersymmetry consistency constraints to identify the correct
ultra-violet cut-offs for the effective low energy theory.  Our results
have a simple interpretation in terms of two-loop unification of
gauge coupling constants at the string scale.
\end{quote}
\begin{flushleft}
\rule{5.1 in}{.007 in}
\end{flushleft}\vspace{-1cm}
$^{\star}${\small Supported in part by the Director, Office \vspace{-4mm}
of Energy Research, Office of High Energy and \linebreak
Nuclear Physics, Division of High \vspace{-4mm}
Energy     Physics  of  the  U.S.    Department
of   Energy   under   Contract
DE-AC03-76SF00098  and in part by the  National\vspace{-4mm}
Science  Foundation under grant
PHY--90--21139.}\\[-2mm]
$^{\dag}${\small Supported in part by the Northeastern University
\vspace{-4mm}
Research and Development Fund and in part by the National Science
Foundation under grant PHY--91--07809.}\\
January 1992
\end{titlepage}

\section{Introduction}\indent

Over the past few years there has been considerable progress
in understanding the structure of effective actions
describing the physics of massless fields in four-dimensional
superstring theory. The basic strategy \cite{dixon}
that has proven to be very successful has been to
extract the relevant terms in the field-theoretical Lagrangian from
the S-matrix elements computed within the full-fledged superstring
theory. In this way many important quantities have been determined
at the classical level, including the
K\"ahler potentials and the gauge and Yukawa couplings for
orbifold compactifications of heterotic superstrings.

Another important development was the observation that the duality
symmetry \cite{dual}
between small and large radius toroidal compactifications
extends to a much larger symmetry group of the so-called target
space modular transformations acting on the moduli fields \cite{sw}.
This symmetry can be very helpful when studying the
moduli-dependence of the effective actions for orbifold compactifications
\cite{flt}.

More recently, the program of reconstructing effective Lagrangians
from string amplitudes has been pursued beyond the classical
level, to higher genus in the string loop expansion. In particular,
the moduli-dependence of the one-loop corrections to gauge
couplings has been determined \cite{dkl,ant} by computing the relevant
string-theoretical amplitudes. These string loop corrections
turn out to be manifestly
invariant under the modular symmetry transformations.
The results of \cite{dkl,ant} provide important constraints
on the form of the effective action describing the physics
of massless string excitations. This is due to the fact
\cite{louis,dfkz,ovrut}
that the radiative corrections generated by quantum
loops of massless particles violate the target space modular invariance.
Since string theory is modular invariant to all loop order
\cite{giveon}, the effective field theory action must contain some
counterterms that cancel this ``modular anomaly'' in a way analogous to
the Green-Schwarz mechanism \cite{gs}. In this paper we determine the
appropriate combination of counterterms
by matching the field-theoretical
couplings of axionic moduli to the gauge bosons
with the corresponding string-theoretical couplings. In addition,
supersymmetry constraints permit us to identify the correct
ultra-violet cut-offs for a consistent Pauli-Villars regularization
of the effective low energy theory.

Two classes of axions will be particularly important for our discussion:
the universal axion and the model-dependent axionic moduli.
The universal dilaton and axion are manifestations of the same
superstring excitation that creates the graviton;
hence they are present in all possible compactifications
of the heterotic superstring theory. The axion
corresponds to a two-index (Minkowski) antisymmetric field $b_{\mu\nu}$, and
together with the dilaton and dilatino it forms one linear
multiplet $L$ \cite{fz} of supersymmetry. The effective tree-level
supergravity Lagrangian for the dilaton sector is readily obtained \cite{wit}
from the ten-dimensional Lagrangian
by dimensional reduction, followed by
the so-called duality transformation on the axion,
from the two-index antisymmetric form to a pseudoscalar field.
The dilaton supermultiplet is represented then by a scalar chiral
superfield which is usually denoted by $S$.
This allows a natural incorporation
of the dilaton and axion into the K\"ahler
structure of the locally supersymmetric four-dimensional sigma model
describing the massless excitations of the compactified superstring.
At the classical level, the K\"ahler potential has the form:
\begin{equation}
K=G-\ln(S+\sbar)\, ,
\label{Ktree}\end{equation}
where $G$ is an $S$-independent
function of all other chiral superfields.  The vacuum expectation
value (VEV) of the dilaton\footnote{We use upper case Roman
and Greek letters for chiral supermultiplets and the
corresponding lower case
letters for their complex scalar components.}
$s$ determines the gauge coupling constant $g$:
\begin{equation}
\frac{1}{g^2} = \langle {\rm Re}\, s\rangle
\label{gbare}\end{equation}
at the string scale $\mst$ which is related to the Planck scale
$\mpl$ by $\mst = g\mpl$ \cite{wit}.

The model-dependent axions correspond to the pseudoscalar components
of the chiral moduli superfields $T$
whose vacuum expectation values determine the geometry
(metric tensor) of the compactified dimensions.
The symmetry group of modular transformations depends on the
particular orbifold background, however it always contains at least
one $SL(2,\bfZ)$ subgroup which acts on a generic modulus $T$ as:
\begin{equation}
T\rightarrow\frac{aT-ib}{icT+d}~,~~~~~ad-bc=1~,~~~~~~a,b,c,d\in\bfZ~.
\label{mod}\end{equation}
The orbifolds that have been discussed most extensively
in the context of modular symmetries are the orbifolds with gauge
group $E_8\otimes E_6\otimes U(1)^2$ \cite{dixon,louis}. They contain
three untwisted
(1,1) moduli $T^I,~I=1,2,3$, which transform under $SL(2,\bfZ )$
as in eq.(\ref{mod}). In order to make our
discussion as explicit as possible, we consider here this one particular
class of orbifolds.
The corresponding K\"ahler potential is:
\begin{equation}
G=\sum_{I}g^I+\sum_{A}
\exp(\sum_{I}q^I_Ag^I)|\Phi^A|^2
+O(\Phi^4)\, ,
\label{gtree}\end{equation}
where $g^I=-\ln (T+\tbar)^I$, and the exponents $q^I_A$ depend
on the particular matter field $\Phi^A$ as well as on the modulus
$T^I$ in question \cite{louis}. Our considerations can be generalized
in a straightforward way to other orbifold models.

The transformation~(\ref{mod}), supplemented by the appropriate
transformations on the matter superfields $\Phi^A$ and chiral rotations
on the gauginos $\lambda$:
\begin{equation}
\Phi^A\to\exp (-\sum_Iq^I_AF^I)\,\Phi^A ,~~~~~~~~
\lambda\rightarrow e^{-\frac{1}{4}(F-\bar{F})}\lambda \, ,
\label{chiral}\end{equation}
where
\begin{equation}
F=\sum_IF^I=\sum_{I}\ln(icT^I+d)\, ,
\label{funF}\end{equation}
effects a K\"ahler symmetry transformation
\begin{equation}
K\rightarrow K+F+\bar{F}
\label{Ktran}\end{equation}
in the tree-level supergravity Lagrangian. The tree-level Lagrangian is
invariant under such transformations.

It has been shown before by other authors \cite{louis,dfkz,ovrut} that
the one-loop effective Lagrangian $\lag_{\rm 1-loop}$
computed within the framework of the effective field theory
of massless string excitations is {\em not\/} invariant under the
transformations (\ref{mod}, \ref{chiral}) because of chiral
and conformal anomalies. The variation of the one loop
Lagrangian \cite{louis,dfkz} under the modular transformation (\ref{mod}) is:
\begin{equation}
\delta{\lag_{\rm 1-loop}}=\frac{1}{16\pi^2}\frac{1}{2}\sum_a\sum_I
\alpha^I_a\int d^2\theta (W^{\alpha}W_{\alpha})^a
F^I+{\rm h.c.}.
\label{variation}\end{equation}
Here, the summation extends over the indices $a$ numbering the simple
subgroups of the full gauge group, and the moduli indices $I$.
The constants $\alpha^I_a$ will be specified
in Section 3, where we derive $\L_{1-loop}$.

Since it is known that the full-fledged string theory is invariant
under the orbifold modular transformations (\ref{mod})
to all orders of its loop expansion~\cite{giveon}, the massless
truncation of string theory is inconsistent unless
the effective field theory is supplemented by counterterms
whose variation cancels the modular anomaly.
These counterterms can be interpreted as the result of
integrating out massive fields,
such as Kaluza-Klein excitations and the winding modes.

Two types of counterterms have been discussed in the literature
in the context of modular anomaly cancellation. The first type
\cite{dkl,fmtv} is a simple gauge ``kinetic'' term:
\begin{equation}
{\cal L}_f=\frac{1}{4}\sum_a\int d^2\theta f_a(W^{\alpha}W_{\alpha})^a
+\rm h.c.,
\label{fterm}\end{equation}
where the summation extends over the indices $a$
numbering simple subgroups of the full gauge group,
and the $f_a$ are analytic functions of the chiral superfields.
They transform as:
\begin{equation}
f_a\rightarrow f_a+ \sum_I\eta_a^IF^I\, ,
\end{equation}
with the constants $\eta_a^I$ adjusted in such a way as to cancel
the modular anomaly, or at least a part of it. The second
type of counterterm \cite{dfkz,ovrut}
is the so-called Green-Schwarz term ${\cal L}_{GS}$
which utilizes the linear multiplet representation
for the dilaton supermultiplet \cite{fv,pierre}.
We will write down ${\cal L}_{GS}$
after reviewing in Section 2 some basic properties of linear multiplets.

In Section 4 we will exploit the linear formalism to relate the
axial couplings given in Section 3 to recent results from string loop
calculations \cite{ant}.
This will allow us to determine the correct choice of counterterm, up to
fluctuations about the vacuum configuration for the effective field theory.
In Section 5 we use supersymmetry consistency constraints to determine
the ultra-violet cut-offs for the effective low energy theory.
We also write down the full one-loop
effective gauge coupling constant, including the contributions
from quantum loops of massless particles as well as from the
counterterms. The result has an interesting interpretation
in terms of the two-loop renormalization group equation.
These results can have interesting implications for threshold
corrections in attempts to extract the string
parameters from data on coupling constant unification.

We should mention at this point that some results of Sections
3 and 4 have already appeared elsewhere.
In particular, the correct combination
of counterterms was identified in \cite{dfkz,ovrut}, although with
little explanation. Since the problem of the moduli-dependence of
effective actions has been obscured in the literature,
which includes a number of incorrect statements,
we address it here in a more systematic way. The novel part
of this paper consists of a detailed discussion of axionic
couplings, the determination of ultra-violet cut-offs and the
implications for two-loop coupling constant unification.

\section{The Linear Multiplet Formalism}
\setcounter{equation}{0}\indent

The general formalism for linear
multiplets has been described in Ref.~\cite{pierre}.
In this section we review the salient properties.
The standard linearity condition
is ${\cal DD}L=\overline{\cal DD}L=0$,
where $\cal D$ and $\overline{\cal D}$ are the supersymmetric
covariant derivatives.
The dilaton $l$ enters as the $\theta=\,\bar{\!\theta}=0$ component of $L$.
For further discussion, it is very convenient
to include in the linear multiplet the Yang-Mills Chern-Simons form
\begin{equation}
\Omega_{\mu\nu\rho}=A_{[\mu}^aF_{\nu\rho]a}-\frac{1}{3}c_{abc}A^a_{\mu}
A^b_{\nu}A^c_{\rho}\, ,
\end{equation}
where $c_{abc}$ are the structure constants of the Yang-Mills group.
This can be done by employing the so-called modified linearity
conditions \cite{pierre}:
\begin{equation}
{\cal DD}L= -\sum_a (\overline{WW})^a~;~~~~\overline{\cal DD}L=
-\sum_a (WW)^a~.
\label{lin}\end{equation}
As a result, the Chern-Simons form enters together with the field strength
of the axion field
$b_{\mu\nu}$ as the $\theta\,\bar{\!\theta}$ component of $L$:
$[{\cal D}_\alpha,\overline{\cal D}_{\dot{\alpha}}]L|_{
\theta=\,\bar{\!\theta}=0}=-2\sigma_{
\mu\alpha\dot{\alpha}}{\cal H}^\mu$, where
\begin{equation}
{\cal H}^\mu =\frac{1}{2}\epsilon^{\mu\nu\rho\lambda}(\partial_\nu
b_{\rho\lambda}-\frac{1}{2}\Omega_{\nu\rho\lambda})+\frac{1}{2}
\lambda^a\sigma^\mu\bar{\lambda}_a\, .
\label{hmu}\end{equation}
The gauge invariance of a multiplet satisfying the
modified linearity condition is ensured by imposing
the appropriate transformation properties for $b_{\mu\nu}$.

The Lagrangian describing one linear multiplet coupled
to supergravity and matter in the presence of a Yang-Mills
Chern-Simons form is:
\begin{equation}
{\cal L}_{lin}=-3\int d^4\theta EF(Z,\bar{Z},L)\, ,
\label{Llin}\end{equation}
where $F$ is a function of chiral and antichiral
superfields
($Z$ and $\bar{Z}$), and the linear superfield $L$. We work in the
K\"ahler superspace formulation \cite{bggm} in which the supervierbein
$E$ depends implicitly on the K\"ahler potential
$K(Z,\bar{Z}, L)$. In order to exhibit the relation
between the linear and chiral representations for the dilaton
supermultiplet, it is convenient to consider
\begin{equation}
{\cal L}_{lin}=-3\int d^4\theta E\left[ F(Z,\bar{Z}, L)+ \ot
(L+\Omega)(S+\sbar )\right],
\label{linL}\end{equation}
where $L$ is now
an unconstrained superfield and $S$ is a chiral superfield.
Here, $\Omega$ denotes the full Chern-Simons superfield \cite{fv}:
${\cal DD}\Omega=\sum_a(WW)^a$.
The superfield $S$ plays the role of a Lagrange multiplier
whose equations of motion enforce the modified linearity constraint
(\ref{lin}).
After eliminating $S$ by using its classical equations of motion one
arrives at the Lagrangian (\ref{Llin}).

The duality transformation from the linear to the chiral representation
amounts to applying the equations of motion obtained by varying
${\cal L}_{lin}$, eq.(\ref{linL}),
with respect to $L$, in order to express $L$ in terms
of the remaining superfields. After eliminating the linear multiplet,
one obtains the supergravity Lagrangian with the correctly
normalized Einstein term provided that the function $F$ satisfies
the condition:
\begin{equation}
F(Z,\bar{Z},L)+\ot L(S+\sbar)=1\, .
\label{condi}\end{equation}
This condition, combined with the equations of motion for $L$,
yields a differential equation for the function $F$ and
the K\"ahler potential $K$, with the solution
\begin{equation}
F(Z,\bar{Z},L)=1+\ot LV(Z,\bar{Z})+\frac{L}{3}\int
\frac{dL}{L}\frac{\partial K}{\partial L}(Z,\bar{Z},L)\, ,
\label{condii}\end{equation}
where the ``constant of integration''
$V$ is an arbitrary real function of the matter superfields.
Here we will consider the special case in which the K\"ahler potential
has the form:
\begin{equation}
K(Z,\bar{Z},L)=G(Z,\bar{Z})+k(L)\, ,
\label{ksep}\end{equation}
where $G$ is an arbitrary function of chiral matter
superfields. Then the constraints (\ref{condi}, \ref{condii}) give
\begin{equation}
\int\frac{dL}{L}k'
=-[S+\sbar+V(Z,\bar{Z})]\, ,
\label{condiii}\end{equation}
where $k'=dk/dl$.
This equation determines the functional dependence of $L$:
\begin{equation}
L=L(X)~,~~~~~X\equiv S+\sbar+ V(Z,\bar{Z})~.
\label{LX}\end{equation}

When the theory is written in terms of the chiral multiplet $S$,
the Lagrangian reads:
\begin{equation}
{\cal L}_{lin}=-3\int d^4\theta E-\int d^4\theta E\Omega (S+\sbar)\, ,
\label{slin}\end{equation}
The first term is the standard supergravity Lagrangian with
the K\"ahler potential $K=G(Z,\bar{Z})+k[L(X)]$ \cite{dfkz}.
The second term gives rise to a gauge kinetic term \cite{cremmer} of
the form (\ref{fterm}) with a group-independent function $f_a=S$.
This is exactly the tree-level kinetic gauge
term \cite{wit} of superstring supergravity!
In fact, the full tree-level Lagrangian is
given by ${\cal L}_{lin}$ corresponding to the
function $F$ determined from the K\"ahler potential of the form (\ref{ksep}),
with $G$ of eq.(\ref{gtree}),
$k(L)=\ln(L)$, and the ``constant of integration'' $V=0$.
Indeed, eq.(\ref{condiii})
then gives $L=(S+\sbar)^{-1}$; therefore the K\"ahler
potential has the right $S$-dependence.

\section{Anomalies in the Effective Field Theory}
\setcounter{equation}{0}\indent

In this section we evaluate the one-loop corrections that
induce modular anomalies.
The results are obtained from
standard results for the effective field theory
in terms of component fields in
the K\"ahler covariant~\cite{bggm} formulation of
supergravity.\footnote{The commonly used
formulation~\cite{cremmer} of supergravity is obtained from the
K\"ahler covariant one by fixing the
K\"ahler gauge with the choice $F(Z) = \ln W(Z)$
in the K\"ahler transformation (\ref{Ktran}).}

The modular transformations act on fermions as field-dependent
(local) chiral rotations (\ref{chiral}). The invariance of
the tree-level Lagrangian under such transformations, which can
be thought of as a combination of the usual K\"ahler
supergravity transformation and field reparametrizations,
is ensured by the presence of the appropriate fermion connection
in the kinetic energy terms.\footnote{The kinetic energy terms for fermions
are normalized canonically: ${\cal L}^{\rm fermion}_{kin}
=-i\bar{\lambda}\bar{\Dslash}\lambda+\dots$}
The modular anomaly includes both the chiral and
conformal anomalies of ordinary field theories.  The chiral anomaly
arises through
the axial part $A_\rho$ of the (matrix valued, moduli-dependent)
fermion connection. The conformal anomaly
arises \cite{tom,bmk,fmtv,bmk2} through the dependence of
the cut-off(s) $\Lambda$ of the effective supergravity theory on the
moduli fields which transform nontrivially under (\ref{mod});
these cut-offs may be different for different sectors of the theory.

Using standard results,
the one-loop correction to the Yang-Mills Lagrangian
is\footnote{When matter fields
$\ph$ are included in the effective Lagrangian (3.2),
the contributions of chiral supermultiplet loops are
modified~\cite{bgj} by $\Tr_MF^2 \to \Tr_Mf^2,\;
\Tr_MF\tF\to \Tr_Mf\tf$, where $f_{\mu\nu} = [D_\mu,D_\nu]_M =
(F_{\mu\nu})_M + O(\ph)$.}
\begin{equation}
\L_{1-loop} = {1\over 16\pi^2} {1\over 4}[\, 3\Tr_G F^2\ln\Lambda_G^2
-\Tr_M F^2\ln\Lambda_M^2
- 4\Tr(F\tF {1\over \partial^2}\partial^\rho A_\rho)\, ] + O(\ph),
\label{la}\end{equation}
where $\ph$ represents gauge nonsinglet fields,
$F_{\mu\nu} = F_{\mu\nu}^a T_a$ is the (matrix valued) gauge field
strength and Tr$_{G(M)}$ means the trace over the gauge (matter)
representation of the generators $T_a$.
Provided that the axial connection and the cut-offs satisfy the conditions
\begin{equation}
A_\rho^{G(M)} = {i \over 4}(\partial_\rho z^m{\partial\over\partial
z^m}B_{G(M)} - h.c.),~~~~
\;\;\;\; \ln\Lambda_G^2 =\frac{B_G}{3}~,~~~~\ln\Lambda_M^2=-B_M~,
\label{ab}\end{equation}
eq.(\ref{la}) takes the form
\begin{equation}
\L_{1-loop} = {1\over 16\pi^2} {1\over 4}\Tr\left\{[F^2
- i F\tF ]{\partial^\rho
\over\partial^2}\left(\partial_\rho z^m{\partial\over
\partial z^m}B\right)\right\}+ h.c.
\end{equation}
which is contained in the component field expansion of the superspace
result
\cite{grisaru}
\begin{equation}
\L_{1-loop} = {1\over 16\pi^2}
{1\over 8}\int d^4\theta\Tr\left\{W^2{\D^2\over\Box }B\right\}+  h.c.
\label{lasusy}\end{equation}
Since the coefficient of the axial anomaly is unambiguous, the
supersymmetry constraint (\ref{ab}) can be used to determine the correct
ultra-violet cut-offs.  We return to this point in Section 5.
Here we will simply use the fact that the anomalous Lagrangian (\ref{lasusy})
is uniquely determined from the knowledge of the
axial vector couplings.

First consider the gauge sector. For gauginos, the axial part of the
connection is
\begin{equation}
A^G_\mu = {i \over 4}(\partial_\mu z^m{\partial\over\partial
z^m}K  - h.c.).
\label{gconn}\end{equation}
Due to the specific gauge kinetic term, eq.(\ref{fterm})
with $f_a = S$,
the chiral gaugino current $\lambda^a
\sigma^{\mu}\bar{\lambda}_a$ couples also to another axial vector field:
\begin{equation}
A^{\prime \,G}_\mu = {i \over 2}\{\partial_\mu z^m{\partial\over\partial
z^m}\ln (s + \s ) - h.c.\}.
\label{ag}\end{equation}
The corresponding radiative correction of the form (\ref{la})
induces an additional
coupling of ${\rm Im\,}s$ to gauge bosons.
In Section 4 we will
argue that the loop diagrams involving
gauge bosons cancel this radiative correction.
This means that the total gauge sector contribution is of
the form (\ref{la}), with the axial vector field of eq.(\ref{gconn}).

Next consider matter-loop contributions. We will evaluate the
effective Lagrangian (\ref{lasusy}) at vanishing VEVs
for gauge nonsinglet scalar fields.  In
this case only gauge nonsinglet chiral fermion
loops contribute to the chiral anomaly.
The fermion matter connection is
[$\,$here the indices $\alpha=(A,a)$ and $\beta=(B,b)$ label gauge
nonsinglet complex scalar fields $\ph^A_a$ and $\ph^B_b$,
respectively, where $a$ and $b$ are the gauge indices;
the index $m$ runs over all scalar fields $z^m$]:
\begin{equation}
(A_\mu^M)^{\alpha}_{\beta} = -{i \over 4}(\delta_{\beta}^{\alpha}\partial_\mu
z^m{\partial\over\partial
z^m}K - 2\Gamma^{\alpha}_{\beta m}\partial_\mu z^m - h.c.), \;\;\;\;
\Gamma^{\alpha}_{\beta m}\partial_\mu z^m =
K^{\alpha\n}K_{\n \beta m}\partial_\mu z^m,
\label{am}\end{equation}
where the second term assures
invariance under reparame\-tri\-za\-tion of the field variables.

To proceed further, we restrict our attention to the class of
orbifolds discussed in the Introduction, with
the corresponding K\"ahler potential
parametrized by a set of constants $q^I_A$, see eq.(\ref{gtree}).
The following equation, which follows from
eq.(\ref{condiii}), is particularly
useful in evaluating the reparametrization invariance connection
coefficients $\Gamma^{\alpha}_{\beta m}$:
\begin{equation}
K_m = \left\{ {-L~~~~~~~~~~\;{\rm if}~ z^m=s\atop
G_m -L V_m~~{\rm if}~ z^m\ne s.}\right.
\end{equation}
It will become clear in the next section that it is
sufficient to consider a limited class of ``integration
constants'' $V$ which are all equal in the
limit of vanishing VEVs for gauge nonsinglet fields:
\begin{equation}
V=\omega \sum_{I}g^I+\sum_{A}p_A
\exp(\sum_{I}q^I_Ag^I)|\Phi^A|^2
+O(\Phi^4) = \omega\, G + O(\Phi^2) \, ,
\label{omega}\end{equation}
with the constant $\omega$ to be determined from the requirement of
modular anomaly cancellation at the vacuum; the form of the $O(\Phi^2)$ terms
is dictated by modular covariance.
 After evaluating $\Gamma^{\alpha}_{\beta m}$,
eq.(\ref{am}) becomes:
\begin{equation}
(A_\mu^M)^{\alpha}_{\beta} =
-{i \over 4}\delta_{b}^{a}\left\{
\delta_{B}^{A}\partial_\mu z^m
\frac{\partial}{\partial z^m} [K-2\sum_I
q^I_Ag^I -2\ln (1-p_A L)]
-h.c. \right\}
\label{mconn}\end{equation}

Summing the gauge and matter contributions to the effective Lagrangian
(\ref{la}), and using the appropriate cut-offs as determined by
eq.(\ref{ab}), the result can be interpreted as a term in the expansion
of the following superspace expression:
\begin{eqnarray}
\L_{1-loop}\!\!\! &=&\!\!\! {1\over 16\pi^2}{1\over 8}\sum_a
\int d^4\theta (W^{\alpha}W_{\alpha})^a{\D^2\over\Box }
\left\{-\sum_I\alpha^I_ag^I
\right.\nonumber\\    & &\hskip 2cm
+\left.\rule{0cm}{6mm} (C^a_G-C^a_M)k(L)+2\sum_AC^a_A\ln (1-p_A L)\right\}
 +~ h.c.\, ,\vspace*{-9mm}
\label{lfin}\end{eqnarray}
where
\begin{equation}
\alpha^I_a=-C^a_G+\sum_A(1-2q^I_A)C^a_A\, .
\label{alpha}\end{equation}
Here $C^a_G$ denotes the quadratic
Casimir operator in the adjoint representation
of the gauge sub\-group labeled by
$a$ and $C^a_M=\sum_AC_A^a= \sum_A {\rm Tr}(T_a^A)^2$, with $T_a^A$
denoting the gauge group generator for the representation of $\Phi^A$.
It is convenient to rewrite eq.(\ref{alpha}) as
\begin{equation}
\alpha^I_a=-C+b'^I_a\, ,
\label{alphap}\end{equation}
where $C=30$ is the Casimir operator in the adjoint representation
of $E_8$. The constants $b'^I_a$ were determined in \cite{louis}
by using the \kp exponents $q^I_A$ previously computed in \cite{dixon}
from the appropriate string amplitudes. It turns out that $b'^I_a=0$,
unless the modulus $T^I$ corresponds to
an internal plane which is left invariant under some orbifold
group transformations; this may happen only if an $N=2$ supersymmetric
twisted sector is present. This means that for a large class
of orbifolds, including the familiar $Z_3$ and $Z_7$ orbifolds
which contain
$N=1$ and $N=4$ sectors only, the constants $\alpha^I_a=-C$
are gauge group independent. This fact will have important implications
for the structure of anomalous counterterms.

\section{Higher Genus Supergravity}
\setcounter{equation}{0}\indent

In this section, we shall discuss the implications of
recent computations~\cite{ant} of higher genus axion couplings
for the structure of field-theoretical Lagrangians
describing orbifold compactifications beyond the classical
approximation. We will determine the appropriate combination
of counterterms which are necessary in order to restore
the modular invariance broken by
the loop contributions of eq.(\ref{lfin}).

The variation of the one-loop Lagrangian
under the modular transformations (\ref{mod})
can be computed from eq.(\ref{lfin}). The result has already been written
in eq.(\ref{variation}), with the constants
$\alpha^I_a$ of eqs.(\ref{alpha}, \ref{alphap}). Notice that
due to the assumed modular invariance of the linear multiplet $L$,
only the first term in eq.(\ref{lfin}) contributes to this modular anomaly.
We have already discussed in the Introduction one class of
counterterms, eq.(\ref{fterm}). We begin this section by
discussing another type -- the so-called Green-Schwarz counterterm
\cite{dfkz,ovrut}.

In the linear multiplet formulation \cite{pierre},
the Green-Schwarz counterterm
corresponds to the part of ${\cal L}_{lin}$ involving the ``constant
of integration'' $V$, {\em c.f}.\ (\ref{condii}):
\begin{equation}
{\cal L}_{GS}= -\int d^4\theta E LV\, .
\label{lgs}\end{equation}
If the linear multiplet is invariant
under modular transformations while $V$ varies as
\begin{equation}
V\rightarrow    V+h(Z)+\bar{h}(\bar{Z})\, ,\vspace*{-5mm}
\label{vvar}\end{equation}
then
\begin{equation}
\delta{\cal L}_{GS}=-\frac{1}{4}
\sum_a\int d^2\theta (W^{\alpha}W_{\alpha})^ah(Z)+h.c.+\dots,
\label{lgsvar}\end{equation}
where we neglected the terms containing the gravitino field
and the terms vanishing in the flat gravitational background limit.
Consider for instance the function $V$ of eq.(\ref{omega}),
which transforms with $h=\omega\sum_IF^I$. Then the  variation
(\ref{lgsvar}) has a form very similar to eq.(\ref{variation}); therefore
the corresponding Green-Schwarz counterterm may contribute
to the modular anomaly cancellation.
Notice that the assumed invariance of $L$ under modular transformations
requires through (\ref{LX}) that its dual $S$ transforms as
\begin{equation}
S\rightarrow S-h\, ,
\label{svar}\end{equation}
therefore in the chiral formulation it is
the tree level gauge kinetic term [the second term in (\ref{slin})]
which gives rise to the {\em r.h.s}.\ of eq.(\ref{lgsvar}).

In general, the modular anomaly can
be cancelled by some linear combination
of ${\cal L}_f$ (\ref{fterm}) and ${\cal L}_{GS}$ (\ref{lgs}).
The question as to which combination actually does
appear in the counterterm can be answered
by comparing the results of recent string computations
of the axion couplings to gauge bosons with the similar
effective supergravity computations.

We are interested in the couplings of scalar particles to gauge bosons.
The explicit form of the Lagrangian $\L_{lin}$ (\ref{Llin})
expressed in terms of the component fields is rather complicated;
however, only three types of terms will be relevant for further
considerations. These are:
1) scalar kinetic terms and tree-level scalar couplings to
gauge bosons, 2) gauge kinetic terms, and
3) pseudoscalar couplings to the chiral matter currents, which
induce the couplings to gauge bosons via the usual anomaly diagrams
discussed in the previous section.

1) The kinetic energy terms for the scalar fields are:
\begin{equation}
{\cal L}_{kin}= -\frac{k'}{4l}\,\partial^\mu l\,\partial_\mu l
+ \frac{k'}{4l}\,{\cal H}^\mu {\cal H}_{\mu} +
(lV-{G})_{i\,\bar{\!\jmath}}
\,\partial^\mu z^i\partial_\mu \bar{z}^{\,\bar{\!\jmath}}
+\frac{1}{2}{\cal H}^\mu {\cal V}_\mu\, ,
\label{lkin}\end{equation}
where
\begin{equation}
{\cal V}_\mu=-i(V_j\partial_\mu z^j-h.c.).
\label{vmu}\end{equation}
Here, we list all terms involving ${\cal H}^\mu$ that
give rise to the kinetic energy terms for the universal axion.
It turns out that the last term in
(\ref{lkin}) contains also the only tree-level coupling of
axionic moduli to gauge fields, through
the Chern-Simons form in eq.(\ref{hmu}).
In order to extract these couplings, one expands $V$ in the fluctuations
of scalar fields about their vacuum expectation values. Then the integration
by parts yields
\begin{equation}
\frac{1}{2}{\cal H}^\mu {\cal V}_\mu =
\frac{i}{2}(V_jz^j-V_{\,\bar{\!\jmath}}z^{\,\bar{\!\jmath}})\,
\partial_\mu {\cal H}^\mu
= -\frac{i}{8}(V_jz^j-V_{\,\bar{\!\jmath}}z^{\,\bar{\!\jmath}})\,
[F^a_{\mu\nu}\widetilde{F}_a^{\mu\nu}-2\partial_{\mu}
(\lambda^a\sigma^\mu\bar{\lambda}_a)]+\dots\, ,
\label{hv}\end{equation}
where $z$ denotes now the fluctuation of the scalar component of $Z$
and the derivatives are evaluated at the vacuum expectation
values.

2) The gauge kinetic terms are:
\begin{equation}
{\cal L}_{gauge}~=~-\frac{1}{4}\sum_a\Delta_a^{\!lin}\,
(F_{\mu\nu}^aF^{\mu\nu}_a+2i\lambda^a\!\not\!\! D\bar{\lambda}_a
+2i\bar{\lambda}_a\,\bar{\!\not\!\! D}\lambda^a )\, ,
\end{equation}
where
\begin{equation}
\Delta_a^{\!lin}=-\int\frac{k'}{2l}\, dl -\frac{1}{2}V\, .
\label{delta}\end{equation}
Note that, in agreement with eq.(\ref{slin}), the duality
transformation (\ref{condiii}) gives
$\Delta^{\!lin}_a={\rm Re}\, s$.

3) The axionic moduli couple to the chiral matter currents via the
field-dependent fermion connections, as discussed in the previous section.
There exists only one additional source of couplings
involving gauge nonsinglet fermions:
the interaction term (\ref{hv}) contains the coupling of the field-dependent
vector ${\cal V}_\mu$, eq.(\ref{vmu}), to
the chiral gaugino current
$\lambda^a\sigma^\mu\bar{\lambda}_a$.
The vector field ${\cal V}_\mu$ is, in the linear formalism,
the analogue of $A^{\prime\, G}_\mu$ (\ref{ag})
whose coupling
to the gaugino current is induced
in the chiral formulation by the gauge kinetic term.

When discussing the couplings of axionic moduli to gauge bosons
it is convenient to consider the
three-point correlation function of the two gauge
fields $A^{a\mu},~A^{b\nu}$ and one modulus $t^I$, with momenta
$p_1,~p_2$ and $p_3$, respectively:
\begin{equation}
\lv A^{a\mu}(p_1)A^{b\nu}(p_2)\, t^I(p_3)\rv_{C\! P\,{\rm odd}}
{}~\equiv~\delta^{ab}
\epsilon^{\mu\nu\rho\lambda}p_{1\rho}p_{2\lambda}\,\Theta_a^I
(\langle z\rangle,\langle\bar{z}\rangle)\, .
\label{ampl}\end{equation}
In order to compute $\Theta^I_a$ in the effective field theory
one expands the effective Lagrangian in the fluctuations $t^I$
of moduli scalars about their vacuum expectation values.
The term linear in the fluctuations contains then the contribution
\begin{equation}\frac{1}{4}
\sum_I(\Theta_a^It^I+\Theta_a^{\bar{I}}\bar{t}^{\bar{I}})
F^a_{\mu\nu}\widetilde{F}_a^{\mu\nu}\, .
\end{equation}
As we have already mentioned before, the only contribution
to this amplitude from ${\cal L}_{lin}$ (\ref{Llin})
comes from the Green-Schwarz
term (\ref{lgs}) and is given by
\begin{equation}
\Theta_a^I|_{GS}=-\frac{i}{2}\frac{\partial V}{\partial\langle t^I\rangle}\, .
\label{thgs}\end{equation}
For completeness, we write down the analogous contribution from
${\cal L}_f$ (\ref{fterm}):
\begin{equation}
\Theta_a^I|_{f}=\frac{i}{2}
\frac{\partial f_a}{\partial\langle t^I\rangle}\, .
\label{thf}\end{equation}
In a supersymmetric theory
\begin{equation}
\Theta_a^I=i\frac{\partial\Delta_a}{\partial\langle t^I\rangle}~,~~~~~~
\Theta_a^{\bar{I}}=-i\frac{\partial
\Delta_a}{\partial\langle \bar{t}^{\bar{I}}\rangle}~,
\end{equation}
therefore the couplings of axionic moduli are related
to the moduli-dependence of the effective gauge coupling
constants \cite{dkl}. In fact, the simplest way to determine the moduli
dependence of threshold corrections to gauge couplings in superstring
theory is by computing the axionic amplitudes. This allows
circumventing the problem of infrared divergences present in
any direct computation of $\Delta$.

We proceed now to the discussion of higher loop corrections to the couplings
of axionic moduli to gauge bosons. The effective couplings
induced by field-dependent fermion connections have already been computed
in the previous section. The final result has been written in the
form of the Lagrangian $\L_{1-loop}$ of eq.(\ref{lfin}).
By expanding $\L_{1-loop}$ in the fluctuations of the
moduli fields
about their vacuum expectation values (and keeping $L$ fixed at its VEV)
we obtain:
\begin{equation}
\Theta^I_a|_{loops}=-\frac{i}{2}\,(\omega+\beta^I_a)\,
\langle t^I+\bar{t}^{\bar{I}}\rangle^{-1}~;
{}~~~~~~~~\omega=-\frac{C}{8\pi^2}~,~~~\beta^I_a=\frac{b'^I_a}{8\pi^2}.
\label{thloop}\end{equation}
Here we assumed zero vacuum expectation values of all fields
except for the dilaton and the moduli.

As already announced in the previous section, we have excluded from
the Lagrangian (\ref{lfin}) possible contributions of the composite
vector field $A_\mu^{\prime\, G}$ (\ref{ag}).
In the linear formalism, this is
equivalent to neglecting loop corrections to the
interaction term (\ref{hv}) involving the moduli-dependent
vector ${\cal V}_\mu$ (\ref{vmu}). The {\em r.h.s}.\ of eq.(\ref{hv})
contains the divergence of the chiral gaugino current
$\lambda^a\sigma^\mu\bar{\lambda}_a$. This current is
not conserved due to anomalous loop corrections;
its divergence contains a term proportional to
$F^a_{\mu\nu}\widetilde{F}_a^{\mu\nu}$, hence it induces an additional
coupling of moduli to gauge bosons. However in a consistent
supersymmetric theory this contribution must be cancelled by loop
corrections to the divergence of the topological current\footnote{The
loop corrections to this divergence have
been considered before in ref.~\cite{russians}, and are a subject
of endless controversies. This is the reason why we do not enter
into the complicated issue of actual computations, restricting our remarks
to the consistency requirements only.}
$\epsilon^{\mu\nu\rho\lambda}\Omega_{\nu\rho\lambda}$.
The reason is that the coupling under consideration
originates from the divergence
\begin{equation}
\partial_\mu{\cal H}^\mu=-\frac{1}{4}
[F^a_{\mu\nu}\widetilde{F}_a^{\mu\nu}-2\partial_{\mu}
(\lambda^a\sigma^\mu\bar{\lambda}_a)]\, .
\label{dh}\end{equation}
which receives contributions from both the topological (first term)
as well as the gaugino
currents. This equation is a part of the linearity condition (\ref{lin}),
therefore it must not be modified by loop corrections. In other
words, there should be no radiative corrections to the
{\em r.h.s}.\ of eq.(\ref{dh}).
The same combination of the topological and gluino
current divergences  couples
to ${\rm Im} f$ in an $f$-type counterterm (\ref{fterm})
therefore the corresponding
couplings of axionic fields should also remain equal to
their classical values.
This statement allows rephrasing our arguments in the chiral formulation.
The coupling of ${\rm Im}\, s$ induced by the gauge kinetic
term ($f=S$) remains equal to its classical value; therefore
$\L_{1-loop}$
does not depend on $A_\mu^{\prime\, G}$.
It is worth mentioning that if this were not the case, then
the variation (\ref{svar})
$s\rightarrow s-h$ would give additional contributions
to the modular anomaly which could not be cancelled by any simple
counterterm, {\em c.f}.\ eq.(\ref{ag}).
We conclude that in a theory with Green-Schwarz and $f$-type counterterms
the loop corrections to moduli couplings
are due entirely to the anomalous interactions involving
composite fermion connections, which justifies
using the Lagrangian (\ref{lfin}) to derive eq.(\ref{thloop}).
The final result
in the effective field theory is:
\begin{equation}
\Theta_a^I|_{e\! f\! f}=\Theta_a^I|_{GS}+\Theta_a^I|_{f}
+\Theta_a^I|_{loops}\, ,
\label{theff}\end{equation}
with the corresponding contributions of eqs.(\ref{thgs}, \ref{thf},
\ref{thloop}).

The three-point amplitude of eq.(\ref{ampl}) was computed directly
in superstring theory~\cite{ant}, to all orders in the higher
genus expansion. The result is
\begin{equation}
\Theta_a^I|_{string}=-\frac{i}{2}\beta^I_a\left\{
\langle t^I+\bar{t}^{\bar{I}}\rangle^{-1}+
\frac{d\ln\eta^2(i\langle t^i\rangle)}{d\langle t^I\rangle}\right\} ,
\label{thstring}\end{equation}
where $\eta$ is the Dedekind eta function. We can identify now the
counterterms that are necessary in the effective field theory
in order to reproduce the string-theoretical amplitudes,
by requiring $\Theta_a^I|_{e\! f\! f}=\Theta_a^I|_{string}$, {\em c.f}.\
eqs.(\ref{theff}) and (\ref{thstring}).
They are, up to fluctuations of the matter fields about the vacuum,
\begin{equation}
V=\omega \sum_I g^I~~~~~~~~~~{\rm and}~~~~~~~~~~
f_a=- \sum_I\beta^I_a\ln\eta^2(iT^I)~,
\label{count}\end{equation}
modulo possible redefinitions $V\rightarrow V+h+\bar{h},~f_a\rightarrow
f_a+h$, where $h$ is an arbitrary analytic function [{\em c.f}.\
eqs.(\ref{thgs}) and (\ref{thf})].
Such a combination of counterterms also cancels
the modular anomaly, as can be verified by using the transformation
property
$\eta^2\rightarrow \eta^2e^{F^I}$
and eqs.(\ref{variation}, \ref{fterm}, \ref{lgsvar}).

The function $V=\omega \sum_I g^I$
corresponds to a ``minimal'' Green-Schwarz
counterterm in the sense that any function
that gives $\omega \sum_I g^I$ in the zero limit
of all gauge nonsinglet fields (and twisted moduli) {\em and\/} has
the same $SL(2,{\bfZ})$ transformation properties,
provides an acceptable counterterm.
A string computation of axionic vertices in the
presence of non-zero backgrounds of twisted moduli and matter fields
is needed in order to impose any further restrictions on $V$.
The same comment applies to the function $f$.
Eq.(\ref{count}) is the main result of this section. The agreement between
string theory and the
effective field theory is guaranteed
to all orders in loop expansion by the Adler-Bardeen theorem.

To conclude this section we comment on the axionic
amplitudes in the chiral formulation of the effective
theory. We consider first the
case of $\beta^I_a=0$, {\em i.e}.\ when the moduli decouple from the
gauge fields, {\em c.f}.\ eq.(\ref{thstring}).
In the linear formulation,
the kinetic energy Lagrangian (\ref{lkin}) does not contain
the mixing of the universal axion represented by $b_{\mu\nu}$
with the pseudoscalars belonging to chiral supermultiplets.
However when the universal axion is transformed to the chiral
supermultiplet $S$, eq.(\ref{lkin}) reads
\begin{equation}
\L_{kin} =
-\partial_\mu t^I \partial^\mu \bt^{\bar{J}} G_{I\bar{J}}\, (1 -\omega l\,)
\, +\, l'\,\partial_\mu y \partial^\mu \bar{y}+\,\dots ;
{}~~~~~~~~~~y=s+\omega G_It^I\, ,
\label{skin}\end{equation}
with $t^I$ and $s$ denoting the fluctuations about their VEVs;
therefore ${\rm Im}\, s$ mixes with other pseudoscalars.
The asymptotic states are $y$ and the combinations of the
moduli that diagonalize the metric $G_{I\bar{J}}$ evaluated at the vacuum.
These are the moduli which decouple from the gauge fields. In the case
of $\beta^I_a\neq 0$, they couple with the correct normalization,
up to unknown modulus wave function renormalizations
of order $g^2\sim l$. Note that our results do not depend on
the form of (and eventual radiative corrections to) the linear part $k(L)$
of the K\"ahler potential. The classical limit $g^2\rightarrow 0$
dictates though $k(L)\rightarrow \ln (L)$ as $L\rightarrow 0$.
\section{Effective Cut-offs and Gauge Couplings}
\setcounter{equation}{0}\indent

In this section we will use the results of Section 3
to identify the effective ultra-violet cut-offs for the low-energy
theory. This is possible due to the condition (\ref{ab}) which
is necessary for a manifestly supersymmetric
calculation of the conformal and chiral anomalies.
The ultra-violet cut-offs may be different for
different sectors of the  theory.
We begin with the determination of the cut-off for
the untwisted matter sector.


The untwisted matter sector
consist of three charged 27 or
$\overline{27}$ matter multiplets $\Phi_U^{\alpha},~ \alpha
=(A,a),~A=1,2,3$. The relevant part of the
\kp can be obtained by dimensional reduction
from ten-dimensional supergravity. It corresponds to the coefficients
$q^I_A=\delta_A^I$. By comparing eqs.(\ref{ab}) and (\ref{mconn})
we obtain
\begin{equation}
\Lambda_I^2=e^{G-2g^I+k}~;~~~~~~~~~I=1,2,3.
\label{uvi}\end{equation}
This and the following relations are written with the accuracy
of order $O(g^2)$.
In the symmetric case
$T^1=T^2=T^3=R^2\mst^2+\dots\,$, {\em i.e}.\
when the three complex tori have equal radii $R$,
the cut-offs (\ref{uvi}) are equal
\begin{equation}
\Lambda_U^2=e^{G/3+k}=R^{-2}\equiv \Lambda^2\, ,
\label{uvu}\end{equation}
where we used the relation $\mst=g\mpl$
and set $\mpl=1$
as is appropriate with the normalization condition (\ref{condi}).
The compactification scale $\Lambda$ is often identified with the
unification scale. It was first pointed out in \cite{tv} that
in the case of orbifold compactifications the two scales are different;
this fact will reemerge at the end of this section where
we discuss the total sum of all moduli-dependent
corrections to gauge coupling constants that include contributions
from the matter and gauge loops as well as from the counterterms.

The cut-offs for the
twisted matter fields can be determined in exactly
the same way as in the untwisted case.
They depend on the weights $q^I_A$ which are known from
the results of \cite{dkl}. For instance, in
an $N=1$ twisted sector, $q^I_A=q_A$ and
\begin{equation}
\Lambda^2_{T}(q_A)    = e^{G(1-2q_A) + k} .
\label{uvt}\end{equation}

In general, the following property holds for all
matter fields $\Phi^A$. The ultra-violet cut-off corresponds
to the Pauli-Villars regulator mass that one would
obtain by supplementing the theory with the mirror fields $\Phi'_A$,
in the gauge group representations conjugate to $\Phi^A$,
and adding the superpotential
\begin{equation}
W_{PV}=\sum_A\Phi^A\Phi'_A\, .
\label{pv}\end{equation}
Indeed,
\begin{equation}
m^2_A = (K^{A\bar{A}})^2e^K = \exp ({G-2\sum_Iq^I_Ag^I + k})\, ,
\label{mpv}\end{equation}
therefore the mass $m_A$ is equal to
the cut-off as determined from eqs.(\ref{ab}) and (\ref{mconn}).
In fact it has been shown~\cite{casimir} for the simple
model~\cite{wit} that such a supersymmetric regularization
gives consistent results for quadratically divergent one-loop
correction to the full effective Lagrangian in the limit of a
large number of gauge nonsinglet chiral supermultiplets.

Finally, we use eqs.(\ref{ab}) and (\ref{gconn}) to obtain
the ultra-violet cut-off for the Yang-Mills sector. The result is
\begin{equation}
\Lambda^2_G = e^{K/3}=g^{-4/3}\Lambda^2,
\label{lam}\end{equation}
which is in agreement with previous results~\cite{bmk2},~\cite{fmtv}
in a sense that will be made explicit below.

We conclude by discussing the effective
gauge coupling constants and their moduli-dependence.
The cut-off dependent gauge and matter loops (\ref{lfin}) combine
with the contributions of counterterms (\ref{count}) to give
a manifestly modular-invariant result\footnote{
We write down the result for a general form
of $V$ (\ref{omega}) with $p_A\neq 0$.}
\begin{equation}
\frac{1}{g^2_a} ~=~ \Delta_a ~=~ \delta_a(l)
-\frac{1}{2}\sum_I\beta^I_a\ln\left[\,\eta^2(it^I)\,
\overline{\eta^2(it^I)}\, (t+\bar{t})^I\,\right] ,
\label{deltat}\end{equation}
where
\begin{equation}
\delta_a(l)=-\int\frac{k'}{2l}\, dl
-\frac{1}{16\pi^2}(C^a_G-C^a_M)k(l)-\frac{2}{16\pi^2}
\sum_AC^a_A\ln (1-p_A l)\, .
\label{deltal}\end{equation}
Here, the {\em r.h.s}. are evaluated at the vacuum expectation
values of the scalar fields.
The effective gauge coupling constants
exhibit nontrivial dependence \cite{dkl,ant} on the moduli only if
the orbifold contains an $N=2$ supersymmetric twisted sector,
{\em i.e}. when some $\beta^I_a\neq 0$.
In particular, these couplings do {\em not\/} depend on the compactification
scale in the case of $Z_3$ orbifold,
in agreement with the general arguments of \cite{tv}.

We close this section with a simple observation
which may be important for the future discussions of
the unification scales. In a supersymmetric gauge theory
with massless matter, the following quantity \cite{russians} is
invariant under the
renormalization group transformations, at least in the two leading
orders of the perturbative expansion in the renormalized
coupling constants $g_a(\mu)$:
\begin{equation}
\delta_a[g_a(\mu),\mu]=\frac{1}{g_a^2(\mu)}-\frac{1}{16\pi^2}
(3C_G^a-C_M^a)\ln\mu^2+\frac{2C_G^a}{16\pi^2}\ln g^2_a(\mu)
+\frac{2}{16\pi^2}\sum_AC_A^a\ln Z_A^a(\mu)\, ,
\label{rge}\end{equation}
where $Z_A^a$ are the renormalization factors for the matter fields. The
``renormalized coupling constants'' $g^{-2}_a = \Delta_a$
in (\ref{deltat}) are
scale-independent quantities, determined by the
ef\-fec\-tive cut-offs and the tree
couplings, that are related to the running coupling constants $g_a^{-2}(\mu)$
by additional terms involving the scale $\mu$ which serves as an infrared
regulator. It is very intriguing that
the moduli-independent part (\ref{deltal})
of $\Delta_a$ can be written as
\begin{equation}
\delta_a(l)~=~\delta_a[g_a(\mu) ,\mu]~,
\label{last}\end{equation}
if we impose as a boundary condition the relations $\mst=g_a(\mst)$
in Planck mass units $\mpl=1$.
Then eq.(\ref{last}) holds provided that we identify
\begin{equation}
g_a(\mst ) = 2l~,~~~~~Z_A^a(\mst)=(1-p_A l)^{-1}
\label{id}\end{equation}
and $k(l)=\ln (l)$,
which is consistent with the results of previous sections.\footnote{
Note that such wave function renormalizations
restore the tree-level form of the
kinetic energy terms for the matter fields,
{\em c.f}.\ eq.(\ref{lkin}).}
We should mention that a similar relation
between the one-loop anomalous lagrangians and the
higher order renormalization group equations
has been pointed out before in \cite{bmk2} and \cite{fmtv}, in the
case of a pure $E_8$ Yang-Mills sector with $C_G = C$ and $C_A = 0$.
Evaluating eq.(\ref{last}) for $a = E_8$ and $\mu = \Lambda$,
{\em i.e.}, at the compactification scale defined in (\ref{uvu}), gives
\begin{equation}
\frac{1}{g_{E_8}^{2}(\Lambda)} = {\rm Re}\,s  + {C\over
16\pi^2}\left\{\ln[g_{E_8}^2(\Lambda)]-k\right\}\, ,
\label{llast}\end{equation}
which agrees
with the boundary condition for the two-loop $\beta$-function found in
\cite{bmk2} and \cite{fmtv}.\footnote{
In these references, $k= -\ln(S+\sbar)$ was used for the
dilaton K\"ahler potential.}
In this case, eq.(\ref{llast}) follows immediately from the $g$-dependence
of the gaugino cut-off (\ref{lam}).

The present formalism, which incorporates the constraints of both
modular invariance and results from higher genus string theory, has a natural
interpretation with the string scale $\mst$
as the scale of two-loop unification [{\em c.f}.\ eqs.(\ref{last},
\ref{id})], up to
the possible additional, moduli-dependent threshold corrections in
(\ref{deltat}) for certain orbifolds.
Determining whether or not this interpretation is fully consistent
requires an understanding of the renormalization of the matter fields.

\section{Conclusions}
\setcounter{equation}{0}\indent

The linear formalism for the dilaton supermultiplet
provides a natural framework
for incorporating a Green-Schwarz counterterm that cancels the modular anomaly
induced by quantum corrections in the effective low energy field theory. The
modular anomaly is related to the axial and conformal anomalies of ordinary
field theory through the nontrivial transformation properties of the axial
currents and the cut-offs of the effective theory under modular
transformations.
There are {\it a priori\/}
three types of axial currents coupled to fermions in a
general supergravity theory: the axial $U(1)$ current associated with K\"ahler
invariance, the K\"ahler connection for matter associated with invariance
under field
redefinitions, and an additional axial current coupled to gauginos that arises
from the noncanonical form of the gauge supermultiplet kinetic energy. We have
argued that the contribution of this last current to
the one-loop chiral anomaly
must be cancelled by other contributions involving gauge-sector loops. The
remaining axial currents are determined once the K\"ahler potential is
specified. In effective field theories from orbifold compactifications of
superstring theory the K\"ahler potential is known at the classical level to
lowest order in
fluctuations about the vacuum; we have studied the induced anomalous
couplings in the same approximation.

    Results from string theory imply that the fully quantum corrected theory is
modular invariant to all loop orders.  This implies that the effective low
energy theory must be modified to include anomaly cancelling counterterms. The
anomalous terms induced at one loop of the unmodified effective field theory
include couplings of the Yang-Mills field
to the axionic moduli; these couplings
are changed once the counterterms are included, and the full Im$\, tF\!\tF$
vertex depends on the
precise choice of counterterm.  As discussed in the text, there
are different types of counterterms that can by used to cancel the modular
anomaly.  We have used the results of string loop calculations of the
axion-Yang-Mills vertex to determine the correct choice of counterterm.
For those cases in which this coupling has been shown to vanish, implementation
of anomaly cancellation is particularly straightforward within the linear
formalism, because, in contrast with the chiral dilaton multiplet $S$,
the linear dilaton multiplet $L$ is modular invariant and unmixed with the
moduli multiplets to all orders in the gauge coupling constant.

    The forms of the chiral and conformal anomalies are related by
supersymmetry.  We used this constraint to determine
the field dependence of the
cut-offs of the effective low energy theory which appear explicitly in
the component field expression for the conformal anomaly.  The results for
matter fields coincide with the regulator masses obtained by
a supersymmetric (but not modular invariant) Pauli-Villars regularization
via the introduction of a bilinear superpotential for heavy regulator fields,
and the result for the gauge sector agrees with results from earlier studies of
effective lagrangians for gaugino condensation.  Our final result for the
renormalized gauge coupling constants has a simple interpretation in terms of
the renormalization group invariants,
with two-loop unification occuring at the string
scale (up to possible additional moduli-dependent threshold effects that can be
present in orbifold compactifications with an $N=2$ twisted sector).

    Our results should have interesting implications for applications of the
renormalization group equations to the low energy theory, such as threshold
corrections to coupling constant unification and effective lagrangians for
hidden gaugino condensation.
\vskip 1cm

\noindent{\bf Acknowledgement}
\vskip 12pt
We wish to thank Ignatios Antoniadis,
Jan Louis, Vadim Kaplunovsky, and especially Pierre Bin\'etruy
for many helpful discussions.
T.R.T. acknowledges the hospitality of Stanford Linear Accelerator Center
and Lawrence Berkeley Laboratory, where this work began.
This work  was  supported  in  part by the
Director, Office of Energy Research, Office of High Energy and Nuclear Physics,
Division of High Energy Physics of the U.S. Department of Energy under Contract
DE-AC03-76SF00098, in  part by the National  Science Foundation under grants
PHY-90-21139 and PHY-91-07809 and in part by the Northeastern University
Research and Development Fund.

\end{document}